\documentclass[
    ,final            
  ]
  {aipproc}
\layoutstyle{6x9}

\newcommand\ion[2]{#1$\;${\small{#2}}\relax}%
\newcommand {\Lya}    {Ly$\alpha$}   
\newcommand {\HI}     {\ion{H}{I}}   

\newcommand {\kms}    {km~s$^{-1}$}

\newcommand {\etal}  {et~al.}

\begin{document}

\title[COS and Future Missions]{The Cosmic Origins Spectrograph and 
    the Future of Ultraviolet Astronomy}


\author{J. Michael Shull}{address={University of Colorado, 
CASA, Department of Astrophysical \& Planetary Sciences, 
391-UCB, Boulder, CO 80309}}

\begin{abstract}
I describe the capabilities of the {\it Cosmic Origins Spectrograph}, 
scheduled for May 2009 installation on the {\it Hubble Space Telescope}.  
With a factor-of-ten increase in far-UV throughput for moderate-resolution 
spectroscopy, COS will enable a range of scientific programs that study 
hot stars, AGN, and gas in the interstellar medium, intergalactic medium,
and galactic halos. We also plan a large-scale HST Spectroscopic 
Legacy Project for QSO absorption lines, galactic halos, and AGN outflows.  
Studies of next-generation telescopes for UV/O astronomy are now 
underway, including small, medium, and large missions to fill the imminent
ten-year gap between the end of {\it Hubble} and a plausible launch of 
the next large mission.  Selecting a strategy for achieving these goals 
will involve hard choices and tradeoffs in aperture, wavelength, 
and capability.  
\end{abstract}

\maketitle


\section{Introduction}

In the near term, the future of ultraviolet space astronomy is reasonably
bright.  The {\it Galex} satellite was well-ranked in NASA's recent
senior review and will continue its programs of broad-band spectral 
imaging in the ultraviolet.  For UV spectroscopy, the Space Shuttle is 
scheduled to install the {\it Cosmic Origins Spectrograph} (COS) on the 
{\it Hubble Space Telescope} (HST) in mid-2009.  Although delayed, the COS 
far-UV (FUV) spectrograph will provide an order-of-magnitude increase in 
throughput for moderate resolution ($R \approx 20,000$ or 15 \kms) 
spectra between 1150--1775 \AA.

However, beyond the lifetimes of {\it Galex} and HST/COS, there may be 
a hiatus in ultraviolet-optical (UV/O) space astronomy.  Following 
a successful Hubble Servicing Mission 4 (May 2009), we anticipate 
an additional 5--10 years of {\it Hubble} lifetime.  However, after 
mission completion of the two Great Observatories, {\it Hubble} 
and {\it Chandra}, we might expect a gap of perhaps 10 years in both 
UV/O and X-ray space astronomy until after 2025, when major new missions 
could be approved and built.  
In this talk, I will briefly describe some of the scientific projects 
enabled by COS, then turn to strategies for filling this 10-year 
gap with new missions in the UV/O spectral bands.  Some key issues are:
\begin{itemize} 

\item What should be the science drivers for new UV/O missions?

\item What is the correct distribution of small, medium, and large missions?

\item How should the UV/O community advocate the best science? 

\item What are tradeoffs in aperture, wavelength band, and other capabilities?

\end{itemize} 

\section{Major Science Goals for COS}

The {\it Cosmic Origins Spectrograph} (COS) was designed to provide
high-throughput ($A_{\rm eff} > 2000$ cm$^2$), moderate-resolution  
($R \approx 20,000$) point-source spectroscopy in the far-ultraviolet
($\lambda \geq 1150$~\AA).  This capability would provide 
a factor-of-ten increase in throughput over previous UV spectrographs 
on {\it Hubble} over the range 1150--1775 \AA.  In addition, a  
near-ultraviolet (NUV) channel (1750--3200 \AA) was added, although 
with only a moderate improvement (factors of 2--3) in effective 
area over STIS.  The capabilities and pre-launch performance of COS were 
described by Froning \& Green (2008), excerpted in Table 1 below.  
The proposed science objectives were given by Morse et al. (1998) and
updated briefly in this section. \\


\begin{table}[h] 
 \caption{COS Far-Ultraviolet Wavelength Modes}
 \begin{tabular}{llclc}
\hline
Grating & Wavelength     & Wavelength Range  & Resolution  &  Average
      Sensitivity \\
        & Coverage (\AA) &  (per expo, \AA)  & ($R = \lambda/\Delta \lambda$) &
      cts/s/resl/[erg~cm$^{-2}$~s$^{-1}$~\AA$^{-1}$] \\ 
\hline
G130M   & 1150--1450  & 300     & 20,000--24,000  & $1 \times 10^{13}$  \\
G160M   & 1405--1775  & 375     & 20,000--24,000  & $7 \times 10^{12}$  \\       
G140L   & 1230--2050  & $>$820  & 2400--3500      & $3 \times 10^{13}$  \\       
\hline
\end{tabular}
\end{table}


\noindent
The major science goals of the Guaranteed Time Observations (GTO) with
COS are enabled by the ability to observe faint targets ($V \approx 17.5$) 
at 15 \kms\ resolution in the ultraviolet.  This allows COS to acquire 
spectra of active galactic nuclei (AGN), hot stars (OB stars, 
white dwarfs, cataclysmic variables, binaries) in the Milky Way, and
O~stars in neighboring galaxies.   Planned observations also include 
studies of planets, comets, disks, and extra-solar planet transits. 
Although COS has no long-slit capability, it will be able to perform 
emission-line measurements of nebulae, supernova remnants, 
and starburst galaxies, although with degraded spectral resolution.      

\begin{figure}
  \includegraphics[height=.400\textheight]{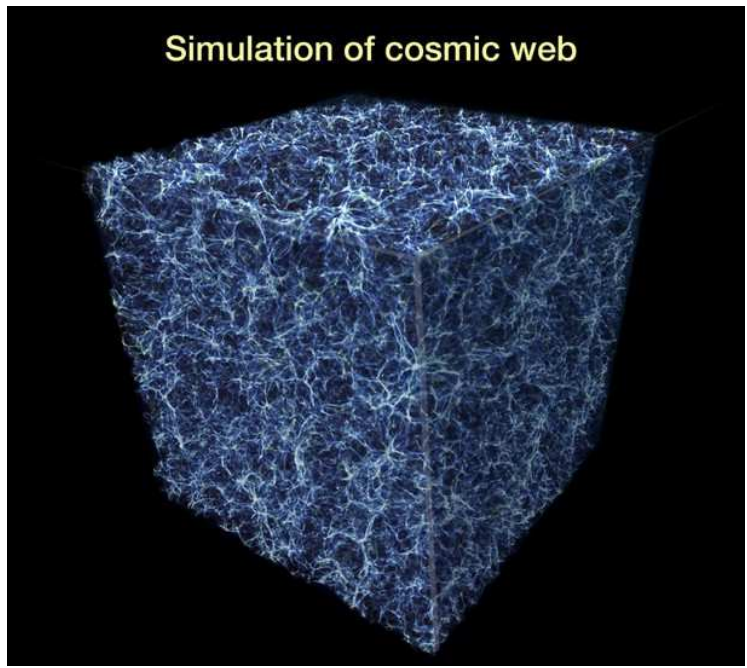}
  \caption{Distribution of baryons in the ``Cosmic Web" at
   $z \approx 0.1$ from IGM simulations (Hallman et al.\ 2007). 
   Large-scale structure develops in filaments, detectable 
   (Danforth \& Shull 2008; Tripp \etal\ 2008) in both
   cold \Lya\ absorbers and in shock-heated gas enriched to 
   1--10\% solar metallicity. }
\end{figure}

Many of the COS science programs involve acquiring absorption-line 
spectra of interstellar and intergalactic gas in the Milky Way and 
its halo, in intervening galaxies, and in the intergalactic medium (IGM).  
Approximately half the 552 GTO orbits are devoted to programs of IGM 
science that focus on high-level questions:
\begin{itemize}

\item How was the ``Cosmic Web" of intergalactic gas formed and  
      structured?   

\item Where are the ``missing baryons" synthesized in the Big Bang?

\item What processes determine the IGM phases in temperature and ionization? 
 
\item What is the chemical composition and spatial distribution of
     metals in the IGM?

\end{itemize} 
The common theme of these studies is to understand the geometry
and composition of the Cosmic Web (Figure 1).  The intergalactic
gas is predicted to consist of complex, multi-phase structures
(Dav\'e \etal\ 2001; Cen \& Ostriker 2006) 
formed by gravitational instability and influenced by feedback  
from star formation in galaxies (ionizing radiation, outflows,
and metal transport).  To conduct these investigations, the COS-GTO 
team formulated a set of observing projects, using 43 background 
AGN as targets to probe the IGM, as well as intervening galaxy halos
and the Milky Way interstellar medium (ISM).  These studies are grouped 
into four projects totalling 253 orbits.  

\noindent
{\it Large-Scale Structure in Baryons (100 orbits, 18 AGN).} 
This program will study the IGM \Lya\ absorbers and associated 
metal-line systems, together with intervening galaxy halos
and Galactic high velocity clouds (HVCs).  Further analysis provides
baryon content, ionization state, covering factor, velocity distribution,
and metallicity. \\ 

\noindent
{\it Warm-Hot Intergalactic Medium (100 orbits, 17 AGN).} 
Absorption studies of highly ionized gas (O~VI, O~V, O~IV, 
N~V, Ne~VIII, etc) and broad \Lya\ lines in redshifted  IGM
absorbers out to $z \approx 0.7$.  Many of these lines have
rest wavelengths in the EUV, and enter the COS band at moderate 
redshifts (e.g., O~VI $\lambda1032,1038$ at $z \geq 0.12$ and Ne~VIII
$\lambda 770, 780$ at $z \geq 0.48$). \\ 

\noindent
{\it Great Wall Tomography (19 orbits, 4 AGN).}
Four background AGN targets will be studied for IGM absorbers in common 
(or absent) to estimate their transverse sizes.  Uncertainty in the 
characteristic size of the \Lya\ absorbers is a systematic error in 
determining their physical density ($n_H$) and baryon content 
($\Omega_{\rm IGM}$). \\  

\noindent
{\it He~II Reionization Epoch (27 orbits, 4 AGN).}
The \Lya\ forest of He~II (304~\AA\ rest frame) enters the COS band 
at $z \geq 2.78$.  We will study He~II (Gunn-Peterson) absorption toward 
two AGN, HE~2347-4342 and HS~1700+6416, studied previously by FUSE, 
but with higher signal-to-noise and velocity resolution of COS. 
We will compare these spectra with \HI\ \Lya\ absorbers, probing
the He~II reionization process at $z \approx$ 3.  \\

The GTO team has also developed several projects to study the 
gas content, ionization state, and metallicity of Galactic HVCs 
in the low Galactic halo. 
These studies will allow us to estimate the HVC contribution to 
the mass-infall rate to the disk and infer their nucleosynthetic 
origin (Collins, Shull, \& Giroux 2004, 2007).

\section{HST/COS Community Legacy Project}

COS should make major discoveries in a range of astronomical areas. 
Because the COS/GTO team will devote only half its time to IGM studies,
a comprehensive study of IGM evolution, galactic halos, and interactions 
of gas and galaxies will require considerably more observing time. 
With COS, astronomers have an opportunity to construct a Quasar 
Absorption Line Key Project, but with a far more powerful spectrograph
than FOS, used for the first {\it Hubble} QSO Absorption-Line Key Project 
(e.g., Weymann \etal\ 1998).  The IGM science can be pushed much farther 
in depth, resolution, and efficiency, using the 10--20 times higher 
FUV throughput and 15 \kms\ spectral resolution of COS, compared to 
230 \kms\ with FOS. 
In concert with galaxy surveys out to $z = 0.3$, we have the
opportunity to study the IGM and its relationship with galaxies on
a scale never before possible. Only at low redshifts will galaxy 
surveys reach to sufficiently low luminosity $(L \approx 0.1L^*)$ 
to explore IGM connections with low-mass galaxies (Stocke \etal\ 2006). \\ 

\noindent
{\it STScI Call for HST Legacy-Project White Papers} \\

\noindent
In 2007, the Space Telescope Science Institute (STScI) issued a call 
for white papers, designed to study large-scale HST Legacy Projects 
that use new instruments (WFC3 and COS) to be installed in 
Servicing Mission 4.  In December 2007, a group of us proposed a 
spectroscopic key project with COS, entitled
{\it Structure and Evolution of the Intergalactic Medium}.
We considered a three-year project of 500--1000 orbits, including 
supplements from Director's Discretionary Time.  The project was
designed to obtain a large number of QSO absorption lines:
over 10,000 \Lya\ absorbers and corresponding metals lines,
using 150-200 AGN as background targets and yielding
moderate-resolution spectra at S/N $\approx 30$.
This proposal was well received upon review at STScI, and we
were encouraged to propose formally for a multi-year program
during HST Cycles 18--20.  In addition to the major study of
the IGM, this program would address a range of topics and
phenomena of broad community interest:
\begin{itemize}

\item Map the IGM large-scale structure at $z \leq 0.3$ compared to galaxy
      distribution
\item Characterize the multi-phase baryon content of cold, warm, and hot IGM 
\item Derive connections between low-$z$ and high-$z$ IGM
\item Measure the spatial transport and redshift-evolution of metallicity
\item Measure the chemical extent of galactic halos and galaxy outflows
\item Probe galactic feedback (energy, radiation, metals) into the IGM
\item Measure HVC mass infall, covering factor, ionization, and metallicity 
\item Study quasar intrinsic absorption lines (AGN outflows, feedback to IGM)

\end{itemize}

\noindent
{\it Scientific Rationale for a COS Legacy Project} \\ 

\noindent
The epoch at $z < 1.5$ marks a time of rapid change in the star-formation
rate over the last 9 Gyr of cosmic time, as galaxies
co-evolved with the IGM.  These interactions include clumpy infall, tidal
flows, and galaxy winds that can pollute the intergalactic space with heavy
elements from star formation.  These feedback processes are essential to
understanding the growth and regulation of galaxy sizes, masses, and
metallicities.  Cosmological simulations take great care to accurately
represent gas flows in and out of galaxies.  The IGM therefore provides a
laboratory for testing the predictions of the thermal phases produced by
large-scale structure formation and feedback.

Even at $z < 0.1$, the majority of baryons still reside in the IGM, not
in collapsed structures (galaxies, groups, clusters contain only 5-10\% of
the baryons).  UV spectra from HST and FUSE have identified approximately
half in the \Lya\ forest (HI) absorbers and shock-heated ($10^5$ to $10^6$~K)
IGM traced by O~VI (1032, 1038 \AA) absorption lines.  Studies of absorption
in H~I, He~II, and heavy elements (multiple ions of C, N, O, Si, Mg, Fe)
allow us to follow the evolution of baryons and metals from $z = 0$ back to
$z = 6$.  

As shown in Figure 2, key diagnostic UV lines are accessible
only to space astronomy (satellite telescopes).
With FUSE and STIS, we have reasonable measurements of the
absorption-line frequency, $dN/dz$, of \Lya, O~VI, and six other 
metal ions at low redshifts: Si~III $\lambda1206$, Si~IV $\lambda1394$, 
C~III $\lambda977$, C~IV $\lambda1549$, etc.) facilitate the 
photoionization corrections needed to derive gas metallicity.
Detecting even hotter IGM gas requires X-ray absorbers (O~VII, O~VIII), 
but may come from studies (Savage \etal\ 2005) of the Ne~VIII doublet
(770, 780 \AA) which shifts into the COS band at $z > 0.5$.  Neon is 
abundant ($\sim20$\% of oxygen), and Ne~VIII is sensitive to gas
at $T \approx 10^6$~K.

COS will be able to observe fainter AGN, thereby increasing the total 
redshift pathlength through the IGM out to $z \sim 0.7$.  This will 
improve the statistics and allows us to probe the evolution of IGM 
phases (warm photoionized, warm-hot collisionally ionized gas).
Access to fainter background targets also provides a higher spatial
frequency of AGN on the sky, constraining the size and shape of \Lya\
absorbers, one of the systematic uncertainties in measuring 
$\Omega_{\rm IGM}$, the baryon content in the IGM.
At redshifts $z < 0.3$, we will be able to compare the large-scale
structure in gas with that surveyed in galaxies (Sloan Survey, 2DF, 6dF).
Obtaining a higher frequency of AGN sightlines on the sky will allow
us to gauge how far metal pollution extends from galaxies.

\begin{figure}
  \includegraphics[height=.400\textheight]{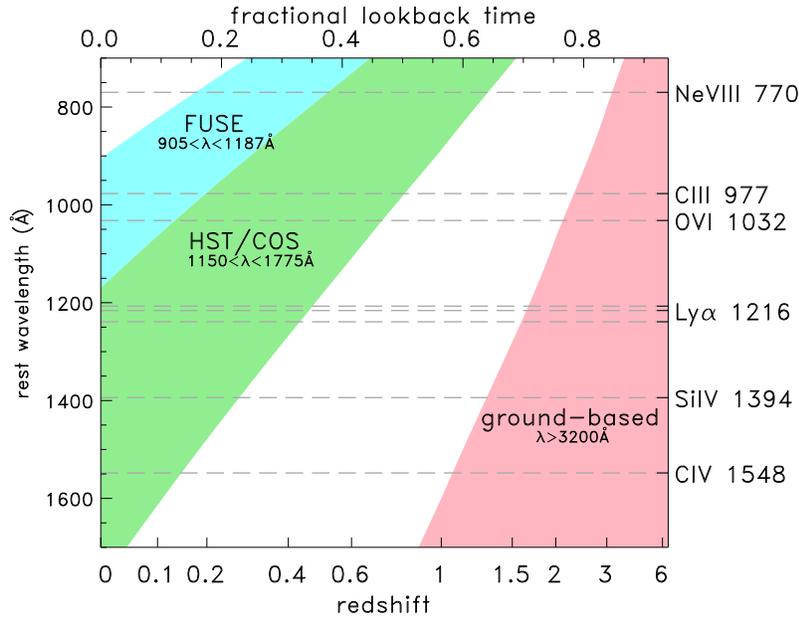}
  \caption{Diagram showing redshifts and lookback times probed
     by various UV lines studied by HST, FUSE, and
     ground-based telescopes (Keck, VLT).  Studying the
    ``near-UV desert" (diagonal white band between $z = 0.5-1.5$)
    poses observational challenges, owing to the lack of
    bright target AGN and to the low quantum efficiency of
    current NUV detectors. }
\end{figure}

COS spectra will have multiple uses, since the sight lines to AGN also 
probe outflows emanating from galactic nuclei, along with the ISM of 
intervening galaxies and Milky Way halo and disk. COS sight lines will 
produce rich spectra of interstellar matter, including infalling HVCs.  
This Legacy Project will be used by observers of ISM, IGM, and galaxy 
formation providing information about QSO continuum emission processes,
nuclear winds, and the broad-line and narrow-line emission regions.  

Other considerations include the usefulness of NUV spectra and  
the efficiency of large FUV wavelength coverage (out to $z = 0.44$ 
in \Lya\ and $0.12 < z < 0.69$ in O VI). We also have the 
opportunity to obtain high-quality spectra (S/N = 30--40) spectra of 
the brightest targets.  COS can observe fainter AGN targets at a
higher spatial frequency, enabling ``IGM tomography" of several 
filaments in the Cosmic Web.  It can probe gas in the outer halos 
of galaxies and HVCs in our own Galaxy.  With greater total 
redshift pathlength ($\Delta z \gg 10$) we may find stronger \Lya\
absorbers (partial Lyman-Limit systems, even a damped \Lya\
absorber).  There are also cosmological applications. With over 
10,000 \Lya\ forest absorbers at $z < 0.4$, we can derive 
power spectra, $P(k)$, in both absorption and transmission.  
Continuing work in the GTO program, we hope to study the evolution 
of the He~II reionization process between $z =2.8-3.2$.

\section{Strategies for the Next-Generation UVO Missions} 

\noindent
As noted earlier, we anticipate an exciting 5--10 years of
UV spectroscopic science with COS and {\it Hubble}.  However,
astronomers are already planning the next generation satellite
missions, in nearly all wavelength bands.  This is prudent,  
given the long lead times and budgetary constraints.  Even
in the best case for NASA astrophysics funding, both UV/O
and X-ray space astronomy are likely looking at a ten-year gap,
extending from perhaps 2016 to 2025.  
{\it Let us therefore advocate to NASA how to fill this gap.} 

First of all, our field of ultraviolet-optical space astronomy
should be proud of what {\it Hubble}, FUSE, and {\it Galex}
have achieved.  There may be some complacency that {\it Hubble} 
could last indefinitely, but it will not.  Astronomers at STScI 
and in the general community are already proposing new ideas for 
missions, including a flagship mission, {\it ATLAS-T} 
(in 8m or 16m varieties) and moderate-cost missions such as 
the {\it Baryon Structure Probe}. NASA has also funded a study 
of a 4-meter mission ({\it THEIA}) for combined wide-field UV imaging, 
UV spectroscopy, and extrasolar planet characterization. 

The major issues facing our discipline include:
a limited ($\sim$\$1B) annual budget for the Astrophysics Division, 
an uncertain set of priorities coming from the 2009-2010 Decadal 
Survey, and many uncompleted projects (JWST, SOFIA, JDEM, IXO, LISA).
Below is a personal list of major science drivers that might
motivate NASA and our astrophysics profession to consider a
UV/O flagship mission, followed by a list of desired capabilities.
Related issues were discussed in the 1999 White Paper on the 
{\it Space Ultra-Violet Optical} Observatory (SUVO) from the 
UVOWG Committee Report:
``The Emergence of the Modern Universe:  Tracing the Cosmic Web"
(http://arxiv.org/abs/astro-ph/9907101). \\ 

\noindent
{\it Top Science goals for a Major UV/Optical Mission} 

\begin{itemize}
\item Map the cosmic web and intervening galactic halos (AGN to 
   $V = 20-21^{\rm m}$)  
\item Study galaxy assembly and stellar populations with deep UV/O imaging 
\item Perform precision cosmology to 1\% (Cepheids, weak lensing) 
\item Characterize extra-solar planets (images, spectra, occulter)
\end{itemize} 

\noindent
{\it What are the Desired UV Capabilities?} 

\begin{itemize}
\item High sensitivity (large $A_{\rm eff}$ spectrograph in the FUV)
\item Good spectral resolution (10 \kms\ with 3 \kms\ mode desirable)  
\item High spatial frequency of targets on the sky (AGN at 1 per arcmin) 
\item Wide-field imaging (10 arcmin and $>20$ arcmin if possible) 
\item Wide-field emission-line mapping (to 10 photon line-units) 
\item Ability to conduct time-domain studies 
\end{itemize}

\noindent
{\bf A Balanced Suite of Missions and Launch Intervals for NASA} \\

Looking forward to the next 20 years, what are the best strategies
for achieving some of the science drivers afforded by UV/Optical
telescopes?   Foremost in the scientific argument are the intrinsic
UV capabilities.  As a result of atomic physics (cross sections, 
resonance line energies, physical diagnostics), 
the ultraviolet band provides the most sensitive 
probes of matter in the form of atoms, ions, and molecules (H$_2$,
HD, CO, etc.)  For instance, the \Lya\ line of hydrogen is $10^6$
times more sensitive to column density, N$_{\rm HI}$, than 21-cm
emission.  UV resonance lines of important ion stages of heavy elements 
(C, N, O, Si, S, Fe) fall in the far-UV band.  
However, other wavelength bands all have strong rationales for how
they are well-suited as astrophysical probes owing to dust-obscuration,
redshifting, or high-energy processes.   So, what is our best course
of action?  
First, we need to come together on
the scientific rationale for a large-aperture, cost-effective 
UV/O space telescope to do the science that is still beyond the
capabilities of {\it Hubble}.  This study is now underway for
{\it ATLAS-T} (see Postman paper).  For the intermediate term, 
we should consider what science can be done with smaller missions,
with focused tasks and conscious tradeoffs:
\begin{itemize}
\item Combine UV spectroscopy with planet-finding (or have dedicated mission?)
\item Separate mission for faint emission-line mapping?  
\item Separate FUSE band ($\lambda < 1150$~\AA) from STIS/COS band
    ($\lambda > 1150$ \AA)? 
\end{itemize}

We also need to be reasonable about cost, if we wish to have regular 
access to space.  This probably means only one astrophysics flagship 
mission per decade. Table~2 illustrates a balanced program of missions 
(small, medium, large) with
costs and launch intervals that fit under the current NASA-Astrophysics
cost umbrella.  Averaged over a decade, they total \$760M/year, 
allowing additional funds for mission operations, data analysis,
technology development, and other important programs (such as theory 
and lab astrophysics).


\begin{table}[h]
 \caption{A Balanced Spaceflight Program}
 \begin{tabular}{lrrr}
\hline
    Mission & Total Cost & Interval & Annual Cost \\
\hline
    {\rm Flagship} &  \$4~B   &  10~yrs  &  \$400~M/yr  \\
    {\rm Probe}    &  \$750~M &   5~yrs  &  \$150~M/yr  \\
    {\rm MidEx}    &  \$360~M &   3~yrs  &  \$120~M/yr  \\
    {\rm SmEx}     &  \$180~M &   2~yrs  &  \$90~M/yr  \\
\hline
    {\rm Total}    &          &          &  \$760~M/yr
  \end{tabular}
\end{table}

However, NASA rarely operates with long-range budgets, as
they are constantly facing cost overruns and budget shortfalls. 
One of the important tasks for us, and for NASA administrators,
is to convince everyone -- the public, our profession, and Congress -- 
that space astrophysics is an exciting area in which to invest.   
The problems are compelling, they excite the public, they stimulate
education and technology, and they deserve more than the \$1B/year
currently devoted to the Astrophysics Division of NASA.  A rising
tide will raise all ships (and many wavelength bands).


\begin{theacknowledgments}
Travel funding to Annapolis was generously provided by NASA/GSFC. 
Our group's research support at the University of Colorado for 
UV studies of the IGM and Galactic halo gas comes from COS grant 
NNX08-AC14G and STScI spectroscopic archive grants (AR-10645.02-A 
and AR-11773.01-A).  I would also like to express appreciation for
many years of NASA support from the FUSE data analysis program,
which enabled so many new ISM and IGM discoveries through 
far-UV spectroscopy.  
  
\end{theacknowledgments}

\bibliographystyle{aipprocl} 

\begin{thebibliography}{9}

\bibitem[Cen \& Ostriker(2006)]{CenOstriker06} 
Cen, R., \& Ostriker, J. P. 2006, ApJ, 650, 560 

\bibitem[Collins, Shull, \& Giroux(2004)]{Collins04} 
Collins, J. A., Shull, J. M., \& Giroux, M. A 2004, ApJ, 605, 216

\bibitem[Collins, Shull, \& Giroux(2007)]{Collins07} 
Collins, J. A., Shull, J. M., \& Giroux, M. A 2007, ApJ, 657, 271 

\bibitem[Danforth \etal(2008)]{DS08}
Danforth, C. W., \& Shull, J. M. 2008, ApJ, 679, 194 

\bibitem[Dave \etal(2001)]{Dave01}
Dav\'e, R., \etal\ 2001, ApJ, 552, 473  

\bibitem[Froning \& Green (2008)]{FroningGreen08}
Froning, C. S., \& Green, J. C. 2008, ApSS, on-line 

\bibitem[Hallman \etal(2007)]{Hallman07} 
Hallman, E. J., \etal\ 2007, ApJ, 671, 27

\bibitem[Morse etal(1998)]{Morse98} 
Morse, J., et al.\ 1998, SPIE, 3356, 361     

\bibitem[Savage \etal(2005)]{Savage05} 
Savage, B. D., \etal\ 2005, ApJ, 626, 776 
  
\bibitem[Stocke \etal(2006)]{Stocke06} 
Stocke, J. T., \etal\ 2006, ApJ, 641, 217  

\bibitem[Tripp \etal(2008)]{Tripp08} 
Tripp, T. M., \etal\ 2008, ApJS, 177, 39 

\bibitem[Weymann \etal(1998)]{Weymann98}
Weymann, R. J., \etal\ 1998, ApJ, 506, 1 

\end{thebibliography}

\end{document}